\documentclass[pra,twocolumn,showpacs,amssymb,amsfonts,amsmath]{revtex4}
\usepackage{graphicx}
\newcommand{\beqa}{\begin{eqnarray}}
\newcommand{\eeqa}{\end{eqnarray}}
\newcommand{\beq}[0]{\begin{equation}}
\newcommand{\eeq}[0]{\end{equation}}
\newcommand{\beqn}[0]{\begin{eqnarray}}
\newcommand{\eeqn}[0]{\end{eqnarray}}
\newcommand{\no}[0]{\\\nonumber}
\newcommand{\np}[0]{{\it e.g. }}
\newcommand{\tzn}[0]{{\it i.e. }}
\newcommand{\proj}[1]{|#1\rangle \langle #1|}
\newcommand{\ket}[1]{|#1\rangle}
\newcommand{\bra}[1]{\langle #1 |}
\newcommand{\tens}[1]{\otimes _{#1=1} ^{k}}
\newcommand{\inner}[2]{ \langle #1 | #2 \rangle}
\newcommand{\melement}[2]{ \langle #1 | #2 | #1 \rangle}
\newcommand{\linia}[0]{\newline\noindent}
\newcommand{\liniaa}[0]{\newline\indent}
\newcommand{\proof}[0]{{\it Proof: }}
\newtheorem{theorem}{Theorem}
\newtheorem{lemma}{Lemma}
\newtheorem{definition}{Definition}
\def\etal{{\it et al.}}
\def\tr{{\rm tr}\; }
\def\cd{{\cal D}}

\def\ca{{\cal A}}

\def\cd{{\cal D}}
\def\ce{{\cal E}}
\def\cp{{\cal P}}
\def\ch{{\cal H}}
\def\eig{\phi_{\rm max}}
\def\LRA{\mathop{-\!\!\!-\!\!\!\longrightarrow}\nolimits}

\begin{document}

\title{Quantum channel capacities - multiparty communication}

\author{Maciej Demianowicz}\email{maciej@mif.pg.gda.pl}
\author{Pawe\l{} Horodecki} \email{pawel@mif.pg.gda.pl}
\address{Faculty of Applied Physics and Mathematics\\
Gda\'nsk University of Technology, 80--952 Gda\'nsk, Poland}

\affiliation{}

\begin{abstract} We analyze different aspects of multiparty
communication over quantum memoryless channels and generalize some
of key results known from bipartite channels to that of multiparty
scenario. In particular, we introduce multiparty versions of
minimal subspace transmission fidelity and entanglement
transmission fidelity. We also provide alternative, local,
versions of fidelities and show their equivalence to the global
ones in context of capacity regions defined. The equivalence of
two different capacity notions with respect to two types of the
fidelities is proven. In analogy to bipartite case it is shown,
via sufficiency of isometric encoding theorem, that additional
classical forward side channel does not increase capacity region
of any quantum channel with $k$ senders and $m$ receivers which
represents a compact unit of general quantum networks theory. The
result proves that recently provided capacity region of multiple
access channel ([M. Horodecki \etal, Nature {\bf 436} 673 (2005)],
[J.Yard \etal, quant-ph/0501045]) is optimal also in the scenario
of additional support of forward classical communication.
\end{abstract}  \maketitle

\section{Introduction}
Theory of quantum channels is nowadays very important domain of
quantum information theory. The case of one sender and one
receiver has been analyzed extensively. For noiseless channel the
coding theorem has been established in \cite{Schu95,Schu96}. The
problem of noisy channels has been defined \cite{BeDiSmoWoo96} on
the basis where minimal fidelity subspace transmission has been
defined. The alternative definition of channel transmission based
on quantum entanglement has been introduced in \cite{BaNieSchu98}
and shown \cite{BaKniNie00} to coincide with that of
\cite{BeDiSmoWoo96}. Moreover it has been shown that forward
classical communication from sender to the receiver does not help
\cite{BaKniNie00}. The research towards channel capacity formula
\cite{Llo97}, \cite{BaSmoTer98}, \cite{BeDiSmo97} has been
finalized by asymptotic coherent information formula in two ways:
there was a proof through conjectured hashing inequality
\cite{HoHoHo00} that has been proven  \cite{DevetakWinter} and  a
direct proof \cite{De05}. The summary of the state of the art of
bipartite zero (one)--way quantum channel capacity can be found in
\cite{KreWer03}. Further other capacity notions measuring the
ability to transmit classical information via quantum channel
\cite{Ho96} or its entanglement assisted analog
\cite{EntanglementAssistedCapacity} have been analyzed. The first
notion leads to well known additivity conjecture (see
\cite{Sho03}) liking four conjectured relations form quantum
information theory. In a mean time much more has been known about
relations between different versions of channel capacities
\cite{BeDeShoSmo04} (see \cite{BeShoScience},\cite{HoHoHoOpp05}).

Some time ago quantum channels with more than one sender/receiver
have attracted more attention. Sending classical information via
multiple access quantum channel has been considered first in
\cite{Winter2001}. The issue of sending quantum information in
general multiparty scenario has been raised in \cite{DuHoCi04}
where in particular it has been shown that quantum broadcast
channel capacity with two-way classical communication is
nonadditive. This is linked to superactivation of multipartite
bound entanglement phenomenon \cite{superactivation}. Recently the
capacity region for multiple access channel have been provided via
quantum-state merging technique \cite{Nature2005},
\cite{NatureExtension} and direct technique using some links
between classical-quantum and quantum-quantum transfer
\cite{YaDeHay05}.

Here we consider most general scenario like in \cite{DuHoCi04} and
generalize most important results from bipartite case. We achieve
it by developing alternative (local) versions  of quantum
fidelities and then adopting with some refinements and
modifications techniques form Refs. \cite{BaKniNie00} and
\cite{BaSmoTer98}. In particular, we consider two versions of
quantum information transfer: minimal subspace fidelity
\cite{BeDiSmoWoo96} and entanglement fidelity from Ref.
\cite{BaKniNie00}. We generalize the result of the latter showing
that the capacity regions defined with respect to both fidelities
coincide under so--called QAEP assumption about the ensembles
describing local senders. We also show that if one drops the
latter assumption then quantum capacity regions of multiple access
channel and $k$-user channel do not change if one removes
encodings which generalizes results of \cite{BaSmoTer98}.

Finally we generalize one of the results of \cite{BaKniNie00}
showing that in general case of $k$ senders and $m$ receivers
forward classical communication does not improve capacity regions.
This result in particular implies  that the regions derived in
\cite{Nature2005}, \cite{YaDeHay05} do not change if we allow the
senders to be supported by one-way classical channel.

The paper is organized as follows. In section II we recall two
definitions of bipartite channel capacities. Then we turn to
multiparty case and introduce two types of fidelities, and define
the corresponding capacities regions. We also provide alternative
(local) versions of the fidelities and show their equivalence to
the previous ones. In section III we show that the capacity
defined under minimal subspace transmission is equivalent to that
defined by entanglement transmission fidelity under assumption of
QAEP of the sources. In section IV we prove, using in particular
local versions of the fidelities, that encoding in zero-way regime
can be always replaced with partial isometry encoding and by its
trace--preserving extension. We consider independently special
cases of $k$-users and multiple access channels  and show that
entanglement transmission capacity regions do not change if we
abandon encoding operations. In section V we show one of the
central results of the paper {\it i.e.} that forward classical
communication does not change the capacity regions for general
$km$-user channel when one can have in general full case of $km$
quantum information ,,flows'' (this is if any of $k$ senders wants
to send quantum information to all the $m$ receivers). The chapter
VI contains the discussion and conclusions.

\section{Preliminaries}
A completely positive operation which is trace--preserving is
called a quantum channel. We will analyze situation in which one
group of parties want to send information to the other group and
particles they send are subject to the act of a channel. In case
of one input--one output we will be talking about single-user
channel (SUC), one-input--m-outputs correspond to broadcast
channel (BC), $k$-inputs--one output is a multiple access channel
(MAC), $k$-inputs--$m$-outputs channel will be called $km$--user
communication channel ($km$--UC, in case of equal $k$ inputs and
outputs just $k$--UC). \liniaa  {\it Remark on notation;-}
Logarithms are taken to basis $2$. We use $\Psi$ to denote
projector $\ket{\Psi}\bra{\Psi}$.

\subsection{Protocols}

Let $C$ denote classical information exchanged between two groups
of parties. We say about zero--way communication when no
information is exchanged, one--way forward or backward if only one
group communicates the other, or two--way when classical messages
fly in both directions. We use following symbols respectively
$C=\emptyset,\rightarrow,\leftarrow,\leftrightarrow$ in these
cases. Let $\textbf{A}=\{\textbf{A}_i\}_{i=1}^k =\{
\{A_i^{\alpha}\}_{\alpha =1}^{l_i}\}_{i=1}^{k}$,
$\textbf{B}=\{\textbf{B}_i\}_{i=1}^m =\{ \{B_i^{\beta}\}_{\beta
=1}^{b_i}\}_{i=1}^{m}$ denote parties taking part in communication
(we have $k$ senders each holding $l_i$ particles and $m$
receivers each getting $b_i$ particles; of course number of
particles on both sides is the same) and $\ch
_{\textbf{A}}=\otimes _{i=1}^k \ch_{\textbf{A}_i}=\otimes _{i=1}^k
\{ \otimes _{\alpha =1}^{l_i}\ch _{\textbf{A}_i}^{\alpha}\}$, $\ch
_{\textbf{B}}=\otimes _{i=1}^m \ch_{\textbf{B}_i}=\otimes _{i=1}^m
\{ \otimes _{\beta =1}^{b_i}\ch _{\textbf{B}_i}^{\beta}\}$ Hilbert
spaces associated to their particles. Let $\varepsilon ^{(n)}$ and
$\mathcal{D} ^{(n)}$ be trace--nonincreasing maps acting as
follows $\varepsilon ^{(n)}: \mathcal{B}(\ch_{\bold{A}}^{\otimes
n})\rightarrow \mathcal{B}(\ch_{iCh}^{\otimes n})$ and
$\mathcal{D} ^{(n)}: \mathcal{B}(\ch_{oCh}^{\otimes n})\rightarrow
\mathcal{B}(\ch_{\bold{B}}^{\otimes n})$, where "i/o Ch" stands
for the input/output channel (in further part of an article we
will call this maps encoding and decoding and set $\dim
\ch_{\bold{A}}=\dim \ch_{\bold{B}}$). We will call a {\it quantum
protocol} (shortly protocol) $\mathcal{P}^C$ supplemented by
classical side information $C$ a set of maps $\{\mathcal{P}
_{n}^{C}\}$ mapping channels $\Lambda ^{\otimes
n}:\mathcal{B}(\ch_{Ch}^{\otimes n}) \rightarrow
\mathcal{B}(\ch_{Ch}^{\otimes n})$ into channels
$\hat{\Lambda}^{(n)}:\mathcal{B}(\ch_{\bold{A}}^{\otimes
n})\rightarrow \mathcal{B}(\ch_{\bold{B}}^{\otimes n})$. Due to
the different usage of classical information we have different
forms of the maps. That is, in case of $C=\emptyset$ we have \beq
\mathcal{P}^{\emptyset}_n(\Lambda^{\otimes n})=\mathcal{D}
^{(n)}\Lambda ^{\otimes n}\varepsilon ^{(n)},\eeq for
$C=\rightarrow$ the map is \beq
\mathcal{P}^{\rightarrow}_n(\Lambda^{\otimes n})=\sum _j
\mathcal{D}_j ^{(n)}\Lambda ^{\otimes n}\varepsilon _j ^{(n)},\eeq
where $\mathcal{D}_j ^{(n)}$ and $\varepsilon _j ^{(n)}$ operate
on the same spaces as $\mathcal{D} ^{(n)}$ and $\varepsilon
^{(n)}$ and $\sum_j \varepsilon _j ^{(n)}=\varepsilon ^{(n)}$. The
most sophisticated case is the two--way communication scenario
$C=\leftrightarrow$ (sometimes called ping--pong protocol)
\cite{Ho03}. Both groups of parties, $\bold{A}$ and $\bold{B}$,
perform POVM-s in turn where any particular POVM can depend on all
of the results obtained in previous ones. This corresponds to
sequence of families of operations $V_{k_1},V_{k_1 k_2}, V_{k_1
k_2 k_3}, ..., V_{k_1 k_2 k_3 k_4 ...k_l}$ with trace preservation
condition $\sum _{k_1 k_2 k_3 k_4 ...k_l}V_{k_1 k_2 k_3 k_4
...k_l}^{\dagger}V_{k_1 k_2 k_3 k_4 ...k_l}=\mathbb{I}$ for all
$l$. Thus, denoting $\bold{k}=\{k_1 k_2 k_3 k_4 ...k_l\}$, we have
\beq \mathcal{P}_n^{\leftrightarrow}(\Lambda^{\otimes
n})=\sum_{\bold{k}}{}^{\bold{B}}
\Lambda^{(n)}_{\bold{k}}\;\Lambda^{\otimes
n}\;{}^{\bold{A}}\Lambda^{(n)}_{\bold{k}}, \eeq where
${}^{\bold{A}} \Lambda^{(n)}_{\bold{k}}$ and
${}^{\bold{B}}\Lambda^{(n)}_{\bold{k}}$ act on the same spaces as
$\varepsilon^{(n)}$ and $\mathcal{D}^{(n)}$ respectively.
\newline
One can show that the latter is the most general form of LOCC
including previous ones. However for the sake of further
convenience we treated these cases separately.

\subsection{Quantum channel rates, fidelities, and capacities
(single user case)}
There are various notions of quantum capacity
rates and transmissions (see \np \cite{KreWer03}). Here we shall
recall two ones that historically  played the most important role.

\subsubsection{Subspace transmission}

We start with the concept of subspace transmission which was
introduced by Bennett \etal  \cite{BeDiSmoWoo96}.
 \liniaa
 The idea is to send pure states from the Hilbert space $H$
being the subspace of a channel input Hilbert space $H_{iCh}$ with
pure state fidelity defined as \beqn
F_s\left(\ch,\Lambda\right)=\min _{\ket{\psi}\in \ch}
\bra{\psi}\Lambda\left( \psi\right)\ket{\psi}.\eeqn We define a
rate of such transmission with a protocol $\cal{P}^C$ as \beqn
R_s({\mathcal P}^C,\Lambda)\equiv
\lim_{n\to\infty}\displaystyle\frac{\log\dim \ch^{(n)}}{n}
\label{subspace}\eeqn and say that is achievable if for a given
protocol we have pure state fidelity tending to one in the limit
of large $n$, \tzn \beq F_s \left(\ch^{(n)},\mathcal{P}
_{n}^{C},\Lambda ^{\otimes n}\right)\stackrel{n\to\infty}{\LRA}
1.\eeq We will say that in this case protocol is reliable. It is
important to note that here we consider only the protocols for
which the limit (\ref{subspace}) exists (in contrast to the
original version where {\it limes supremum } was put in this
place), but it can easily be shown not to be a restriction.
Quantum channel capacity for subspace transmission is defined as
the supremum of all achievable rates produced by all considered
protocols $\mathcal{P}^C$. We use $Q_s^C$ to denote capacity in
this case\beq Q_s^C(\Lambda)=\sup _{\mathcal{P}^C} R_s({\mathcal
P}^C,\Lambda).\eeq
 There is also an alternative way to cope with the problem of
quantum information transmission - entanglement transmission.

\subsubsection{Entanglement transmission}

The idea of entanglement transmission was developed in
\cite{BaKniNie00} and is as follows.
 Alicia produces bipartite state $\Psi _{AB} $ and sends
 its subsystem $B$ down the channel $\Lambda$ to Bob.
The quantity which measures the resemblance of the output
bipartite state to the initial $\Psi_{AB}$ is called entanglement
fidelity and is defined by the relation (we use here notation
staying in agreement with one used in \cite{YaDeHay05}) \beq
F_e(\Psi _{AB},\Lambda)\equiv
\bra{\Psi_{AB}}\mathbb{I}^A\otimes\Lambda^B \left(
\Psi_{AB}\right)\ket{\Psi_{AB}}.\eeq It can be shown that it
depends on $\Psi_{AB}$ only through $\varrho_B$ and $\Lambda$. The
entanglement transmission rate in the given protocol
$\mathcal{P}^C$ for the source $\varrho^{(n)}$ can be defined
formally as \footnote{We slightly modify the previous definition
that involves limes supremum. This does not change the final
rate.} \beqn R_{e}(\mathcal{P}^C,\Lambda) \equiv \displaystyle
\lim _{n\to\infty}\frac{S(\varrho^{(n)})}{n}. \eeqn Here
$\varrho^{(n)}$ is a reduced Alicia's block density matrix
corresponding to bipartite pure state $\Psi_{AB}^{(n)}$ part of
which is transmitted down the new channel $\cp
_{n}^{C}(\Lambda^{\otimes n})$ that involves both encoding and
decoding procedures.

As previously we say that the rate is achievable if\beqn
F_e\left(\Psi_{AB}^{(n)}, \mathcal{P} _{n}^{C},\Lambda ^{\otimes
n}\right) \stackrel{n\to\infty}{\LRA} 1. \eeqn Channel capacity is
defined analogously to the previous case as \beq
Q_e^C(\Lambda)=\sup _{\mathcal{P}^C} R_s({\mathcal
P}^C,\Lambda).\eeq

 Barnum \etal $\;$\cite{BaKniNie00} have shown  that both definitions
coincide and give the same number for sources which satisfy {\it
quantum asymptotic equipartition property} (QAEP) \tzn for all
$\varrho^{(n)}$ that have asymptotically uniform spectrum in a
special sense (see Appendix).

Recently it has also been shown \cite{De05} that it is equal to
the so--called coherent information rate.

\subsection{Quantum channel rates, fidelities, and capacities (multiuser case)}

Let us now turn to the multiuser case and consider MAC, BC, and
$km$--UC.\newline
 We then have in the case of subspace transmission in most general
 case of $km$--UC.
 \beqn &&\hspace{-6mm}F_s\left(\tens{j}\left(\otimes _{\alpha=1}^{l_j}\ch _{j}^{\alpha}\right),\Lambda\right)\equiv
 \min_{\tens{j}\left(\otimes_{\alpha=1}^{l_j}\ket{\psi ^{\alpha}_j}\right)\in
 \tens{j}\left(\otimes _{\alpha=1}^{l_j}\ch
 _{j}^{\alpha}\right)}\\\nonumber&&\hspace{-6mm}
 \left(\tens{j}(\otimes_{\alpha=1}^{l_j}\bra{\psi ^{\alpha}_j})\right)
 \Lambda\left(\tens{j}\left(\otimes_{\alpha=1}^{l_j}{\psi ^{\alpha}_j}\right)
 \right)\left(\tens{j}(\otimes_{\alpha=1}^{l_j}\ket{\psi
^{\alpha}_j})\right)\\\nonumber \eeqn
 and in case of entanglement transmission
 \beqn
&&\hspace{-6mm}F_e\left(\tens{j}\left(\otimes_{i=1}^{l_j}\Psi_{(AB)_{j}}^{i}\right)
,\Lambda\right)\equiv
\left(\tens{j}\left(\otimes_{i=1}^{l_j}\bra{\Psi_{(AB)_{j}}^{i}}\right)\right)\\\nonumber\hspace{-6mm}
&&\mathbb{I}^{\bold{A}}\otimes\Lambda^{\bold{B}}
\left(\tens{j}\left(\otimes_{i=1}^{l_j}\Psi_{(AB)_{j}}^{i}\right)\right)
\left(\tens{j}\left(\otimes_{i=1}^{l_j}\ket{\Psi_{(AB)_{j}}^{i}}\right)\right).\eeqn\
 We will use the term {\it global
fidelities} to name above quantities. In each scenario we can
assign every $i$--th transmission between
 $i$--th (sub)sender (sender sending one of her subsystem) and
 proper receiver in a manner we have done in the single user
case. Literally, we define the rates as (we shall abbreviate
notation here)
 \beq R_{s}^{(i)}\equiv
\lim_{n\to\infty}\displaystyle\frac{\log\dim \ch_i^{(n)}}{n}\eeq
for subspace transmission, where $\ch_i^{(n)}$ is the subspace of
$i$--th input Hilbert space $\ch_{Ch_{i}}^{\otimes n}$ and for
entanglement transmission \beqn R_{e}^{(i)}=\displaystyle\lim
_{n\to\infty}\frac{S(\varrho_{i}^{(n)})}{n},\eeqn  where
$\varrho_{i}^{(n)}$ can be represented as a quantum material
produced by $i$--th source $\Sigma^{(i)}$ but formally is just a
(Alicia's) reduced state of  the bipartite pure state $\Psi
_{(AB)_{i}}^{(n)}$.

Similarly to the single user case we say that the rates are
achievable if there exists protocol for the given type of the
scenario (\tzn for instance it requires product encoding but joint
decoding in case of multiply access channel) such that senders can
reliably, \tzn with global fidelity corresponding to those rates
approaching one, send information to the side of receivers.
Formally we require (following \cite{YaDeHay05}) the global
fidelity to approach unity, but it can be easily seen that this is
equivalent to the same requirement for set of {\it local
fidelities} (see next subsection).

We define quantum channel capacity to be a set of all $L$-tuples
($L=\sum_{i=1}^k l_i$) of achievable rates $[R^{(1)},R^{(2)}, ...,
R^{(L)}]$. Here they will be one of three types: (i) from $k$
senders to one receiver in case of MAC (ii) from $k$ senders to
$m$ receivers in case of $km$-UC and (iii) from single sender to
$k$ receivers in case of BC.
 \liniaa Other capacities will be analyzed in
detail elsewhere \cite{Prep}.

\label{fidelities}\subsection{Alternative expressions for
fidelities}

Instead of global fidelities, expressing resemblance of the whole
state on input to the whole state on the output, we can introduce
local fidelities measuring how each density matrix sent through
the channel was affected by its action. Namely we have for
subspace and entanglement transmission respectively \beqn
&&\hspace{-9mm}F^{(j,l)}_s\left(\ch_j^l,\Lambda\right)=
\min_{\tens{j}\left(\otimes_{\alpha=1}^{l_j}\ket{\psi
^{\alpha}_j}\right)\in
 \tens{j}\left(\otimes _{\alpha=1}^{l_j}\ch
 _{j}^{\alpha}\right)}\\\nonumber && \bra{\psi_j^l}\tr
_{\bold{AB}\setminus(AB)_j^l}
\Lambda\left(\tens{j}\left(\otimes_{\alpha=1}^{l_j}{\psi
^{\alpha}_j}\right)
 \right)\ket{\psi _j^l},\eeqn

\begin{figure}
\centering
\includegraphics[width=9.5cm,height=7cm]{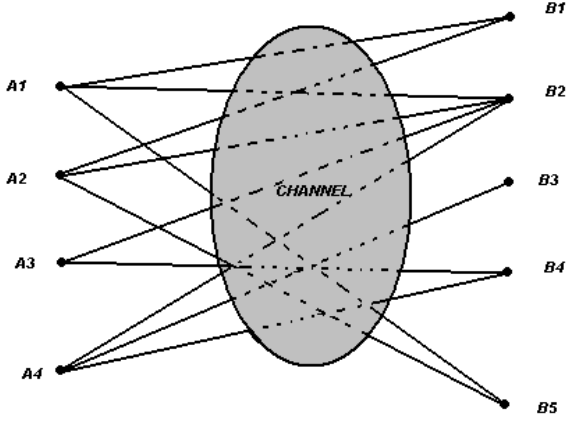}
\caption{General scheme of multiuser communication. An exemplary
45--UC.}
\end{figure}

\beqn &&\hspace{-5mm}F^{(j,l)}_e\left(\Psi
_{(AB)_j^l},\Lambda\right)=\\\nonumber&&\hspace{-5mm}
\bra{\Psi_{(AB)_i^l}}\tr_{\bold{AB}\setminus(AB)_j^l}\mathbb{I}^{\bold{A}}\otimes\Lambda^{\bold{B}}
\left(
\tens{j}(\otimes_{i=1}^{l_j}\Psi_{(AB)_j^i})\right)\ket{\Psi
_{(AB)_i}}.\eeqn

Following lemmas show that we can freely switch from global to
local fidelities (and the opposite way) as they both coincide in
case of high limit.

\begin{lemma}
\label{localfidelities} For any k-partite density matrix
$\varrho_{\bold{k}}$ and arbitrary states $\phi_i$ satisfying
$(\tens{i}\bra{\phi_{i}})\varrho_{\bold{k}}(\tens{i}\ket{\phi_{i}})>1-\epsilon$
($\bold{k}=\{i\}_{i=1}^{k}$) we have (i) for all $l$
$\melement{\phi_l}{\varrho_{l}}>1-\epsilon$ and (ii)
$(\tens{i}\bra{\phi_{i}})(\tens{i}\varrho_{i})(\tens{i}\ket{\phi_{i}})>1-k\epsilon$.
\end{lemma}
\proof Proof is straightforward. Consider property (i). For k=2 we
prove it  by writing partial trace over subsystem $1$ in the bases
$\ket {\phi _1 ^{j}}$ (with $j$ enumerating the basis vectors) and
by taking $\ket {\phi _1 ^{0}}\equiv \ket {\phi_{1}}$ and
considering respective fidelity. We then have \beqn
\bra{\phi_2}\tr _{1}\varrho _{12}\ket{\phi_2}=\sum_{j\ne 0}
\bra{\phi _1^{j}}\otimes\bra{\phi
_{2}}\varrho_{12}\ket{\phi _1^{j}}\otimes\ket{\phi _{2}} +\\
\nonumber
\bra{\phi_{1}}\otimes\bra{\phi_{2}}\varrho_{12}\ket{\phi_{1}}\otimes\ket{\phi_{2}}.\eeqn
which immediately gives the desired bound because of the
positivity of the first term and the assumption.

For any $k>2$ the proof goes in the same way, only instead of
subsystem $2$ we consider all the other subsystems. This gives
property (i) for index $i=1$. By permutation we get the same for
all other indices.

Let us pass to the property (ii). It immediately follows by
multiplication of inequalities from (i) and the fact that
$(1-\epsilon)^{k} \geq 1-k\epsilon $ which easily can be proven by
induction.$\blacksquare$

\begin{lemma} \label{localfidelities-conv}
For any k-partite density matrix $\varrho_{\bold{k}}$ and
arbitrary states $\phi_i$ satisfying
$\melement{\phi_i}{\varrho_{i}}>1-\epsilon_{i}$ we have
$(\tens{i}\bra{\phi_{i}})\varrho_{\bold{k}}(\tens{i}\ket{\phi_{i}})>1-\sum_{i=1}^{k}\epsilon_{i}$.
\end{lemma}

{\it Proof .-} Again here it is enough to prove the observation
for $k=2$ since for higher $k$ the induction method does the job.

As before we introduce orthonormal bases $\ket {\phi _1 ^{j}}$,
$\ket {\phi _2 ^{j}}$. Let us now define $\varrho_{op,rs}\equiv
\langle \phi _1 ^{o} |\langle \phi _2 ^{p} |\varrho_{12} \ket{\phi
_1 ^{r}} \ket{\phi _1 ^{s}}$ . Then we have (because of unit trace
of $\varrho_{12}$)
\begin{equation}
\varrho _{00,00} + \sum_{i\neq 0} \varrho _{i0,i0} + \sum_{j\neq
0}\varrho _{0j,0j} + \sum_{i\neq 0,j\neq 0}\varrho _{ij,ij}=1.
\label{sum1}
\end{equation}
On the other hand the condition
$\melement{\phi_1}{\varrho_{1}}>1-\epsilon_{1}$ takes the form
\begin{equation}
\varrho _{00,00} + \sum_{j\neq 0}\varrho _{0j,0j} >1-\epsilon_{1}
\label{sum2}
\end{equation}
which, if put into the previous inequality, gives $\sum_{i\neq
0}\varrho _{i0,i0} + \sum_{i\neq 0,j\neq 0}\varrho _{ij,ij}<
\epsilon_{1}$ which implies also $\sum_{i\neq 0}\varrho _{i0,i0} <
\epsilon_{1}$. The latter put into the condition
$\melement{\phi_2}{\varrho_{2}}>1-\epsilon_{2}$ rewritten as
$\varrho _{00,00} + \sum_{i\neq 0}\varrho _{i0,i0}
>1-\epsilon_{2}$ leads to $\varrho _{00,00}
>1-\epsilon_{1}-\epsilon_{2}$ which concludes the desired proof
for $k=2$. As mentioned, for higher $k$ the proof goes by
induction.$\blacksquare$\liniaa Application of the above lemmas to
fidelities is obvious, as we can combine both lemmas to get "if
and only if" statement: {\it Global fidelities are high iff local
ones are so.}

\section{Entanglement and subspace
transmission: an equivalence}

In this section we argue that quantum capacity of a quantum
channel for entanglement transmission is equal to that of subspace
transmission in multiuser communication scenarios for sources
satisfying QAEP. What is worth to be stressed equivalence holds
for {\it every} type of the protocol involved \tzn {\it every}
type of classical side information. This due to the fact that the
proof does not specify protocol.

\subsubsection{Subspace transmission follows from entanglement
transmission}

We employ techniques used in \cite{BaKniNie00} and use results of
section IID. Assume each of $k$ senders' every
$\varrho_{B_l}^{\gamma(n)}$ (shortly $\varrho_{l}^{(n)}$) satisfy
QAEP and can be sent reliably (\tzn with high local entanglement
fidelity, consequently global). As noted in \cite{BaKniNie00} we
can then restrict ourselves to $\varrho^{(n)}_l$ projected onto
its typical subspace. We use this fact and assume that this
restricted source is sent with local entanglement fidelity at
least $1-\eta _l$ (by the lemma 2 global entanglement fidelity is
at least $1- \sum _{l=1}^L \eta _l$, $L=\sum_{i=1}^{k}l_i$). We
recursively remove the lowest pure--state fidelity vectors
$\ket{\phi_{l}^{i(n)}}$ (consequently dimensions) from the
$K_{l}$--dimensional support of each $\varrho_{l}^{(n)}$ in such a
manner that for each $m$ we keep an operator
$\varrho_{l}^{(n)}-\sum_{i=1}^{m}q_{l}^{i}\ket{\phi_{l}^{i(n)}}\bra{\phi_{l}^{i(n)}}$
positive (consequently a tensor product of them). We then have
$\varrho_{l}^{(n)}=\sum_{i=1}^{K_{l}}q_{l}^{i}\ket{\phi_{l}^{i(n)}}\bra{\phi_{l}^{i(n)}}$,
which means that $\{q_{l}^{i},\ket{\phi _{l} ^ {i(n)} }\}$
constitute a pure--state ensemble for $\varrho_{l}^{(n)}$. After
our removal procedure we are left (for every $\varrho_{l}^{(n)}$)
with a subspace with dimension $D_l=K_l-n_{(l)}$, where $n_{(l)}$
is the number of removed dimensions, from which each state has
pure--state fidelity \beq \label{fidelity}F_{s}^{(l)}\ge 1 -
\frac{\sum _{p=1}^L \eta _p}{\Pi _{p=1}^{L}\alpha_{p}} , \eeq
($\alpha_p \equiv\sum_{i=1}^{n_{(p)}}q^i_{p}$). We also get the
bound for the dimension of the remaining subspaces. Namely $D_l
\ge (1 - \alpha_l) 2^{n(S_l - \epsilon_l)}$, which gives the rate
for subspace transmission $\frac{\log{D_l}}{n} \ge \frac{\log{(1 -
\alpha_l)}}{n}+ S(\Sigma^{(l)}) - \epsilon_l$. The latter means
that all rates for subspace transmission at least as large as for
entanglement transmission are achievable.
\subsubsection{Entanglement transmission follows from subspace
transmission} We use here a modified version of a well known
theorem \cite{BaKniNie00},\cite{BaSmoTer98}.
\begin{theorem}
If every pure product state $\tens{i}(\otimes_{n=1}^{l_i}
\ket{\Psi^n_i})$ from a space $S=\tens{i}(\otimes_{n=1}^{l_i}
S^n_i)$ have pure--state fidelity $F_{s}(S,\ce)>1-\eta$ then any
density operator $\varrho =\tens{i}(\otimes_{n=1}^{l_i}
\rho^n_i)$, such that $\; Ran (\varrho _i^n)\subseteq S_i^n$, has
entanglement fidelity $F_e(\varrho ,\ce) \ge 1 - O(\eta)$.
\end{theorem}
Proof of the theorem (see \cite{BaKniNie00}) uses two main ideas
(i) not only vector from the bases have pure--state fidelity high
but arbitrary superpositions of them also, and (ii) pure--state
fidelity averaged over phases is high if pure--state fidelity is
high. (Note that we consider here global fifelities which in
virtue of lemmas from section \ref{fidelities} are equivalent to
local ones in context of capacities). To recognize usefulness of
the theorem take uniform density matrices on each $l$--th subspace
$H^{(n)}_{l}$, \tzn $I^{(n)}_{l}/\dim \ch ^{(n)}_{l}$. We conclude
from the theorem that this source can be sent reliably which means
that all rates for entanglement transmission not less than for
that of subspace transmission are achievable. \liniaa We obtained
two opposite inequalities for rates of transmission which means
equality of capacities.

\section{Encodings}

\subsection{Sufficiency of isometric encodings}

As one knows isometric encodings is sufficient to achieve SUC
capacity \cite{BaKniNie00}. We show that it is true for all
classes of quantum channels considered in the paper (SUC, MAC,
$km$--UC). Bearing in mind that the protocol is agreed before
sending any information through the channel so that both encoding
and decoding depend on the source we will construct explicitly
isometry which will serve as an encoding.  We start with recalling
theorem about sufficiency of isometric encodings in case of a
single user channel as its proper application will be a main tool
in proving the upcoming central theorem of the section. We have
\begin{lemma}\cite{BaKniNie00}
Given a trace--non-increasing map $\mathcal{A}$ and a map
$\varepsilon$ trace--preserving on the state $\varrho _B\equiv\tr
_{A}\Psi_{AB}$, for which $F_e (\Psi
_{AB},\mathcal{A}\circ\varepsilon)>1-\eta$, one can always find
such partial isometry $W$ that $F_e (\Psi _{AB},\mathcal{A}\circ
W>1-2\eta$.
\end{lemma}
To apply the above we take $\mathcal{A}=\mathcal{D}\circ\Lambda$,
\tzn concatenation of channel noise and decoding. \linia Central
point of this section is a theorem
\begin{theorem} Given a reliable protocol
$P^{C}=\{\cd^{(n)},\varepsilon^{(n)}\}$,
$\varepsilon^{(n)}=\tens{i}\varepsilon^{(n)}_i$, there always
exists an extendable to a trace--preserving map partial isometry
$W^{(n)}=\tens{i}W_i^{(n)}$ such that a protocol
$\widetilde{P^{C}}=\{\cd ^{(n)}, W^{(n)}\}$ allows for reliable
entanglement transmission with the same rate.
\end{theorem} \proof For clarity we omit a superscript $"(n)"$ in the proof. Using shorthand notation
for $\mathcal{A}$ as previously (the difference is that in the
setting considered now we have $m$--fold tensor product of
decodings) reliable entanglement transmission condition in
multiparty scenario takes the form \beqn\label{reliable}
F_e(\tens{i}\Psi_{(AB)_i},\mathcal{A}\circ\tens{i}\varepsilon_{i})>1-\eta.\eeqn
Assume now $\mathcal{A}$ and $\varepsilon_i$--s have Kraus
representations as follows \beqn
\mathcal{A}(\cdot)=\sum_{\alpha}A_{\alpha}(\cdot)A_{\alpha}^{\dagger},\eeqn
 \beqn
\varepsilon_i
(\cdot)=\sum_{\beta_i}E_i^{\beta_i}(\cdot)E_i^{\beta_i\dagger}.\eeqn
One verifies that independently of what purifications
$\Psi_{(AB)_i}$ we choose we have \footnote{This is a multiparty
generalization of the formula (11) of \cite{BaKniNie00}}
\beqn\label{fidisom}
&&F_e(\tens{i}\Psi_{(AB)_i},\mathcal{A}\circ\tens{i}\varepsilon_{i})\\\nonumber&&=\sum_{\alpha}
\sum_{\beta_1\beta_2...\beta_k}\\\nonumber&&\Big|\sum_{\gamma_1\gamma_2...\gamma_k}\left(\tens{i}\bra{\widetilde{\phi
_{B_i}^{\gamma_i}}}\right)A_{\alpha}\tens{i}\left(E_{i}^{\beta_i}\ket{\widetilde{\phi
_{B_i}^{\gamma_i}}}\right)\Big|^2 ,\\\nonumber\eeqn where
$\phi$--s come from spectral decompositions of $\varrho$--s, \tzn
\beqn \varrho_{B_i}=\sum_{\gamma_i}\lambda_{\gamma_i}\proj{\phi
_{B_i}^{\gamma_i}}\equiv\sum_{\gamma_i}\proj{\widetilde{\phi
_{B_i}^{\gamma_i}}}.\eeqn Now let us define some operators by
partial inner product as follows \beqn\label{krausA1}
A_{\alpha,\beta_2 \beta_3... \beta_k}^{(1)}\equiv\sum_{\gamma_2
\gamma_3 ... \gamma_k}\left(\otimes_{i=2}^{k}\bra{\widetilde{\phi
_{B_i}^{\gamma_i}}}\right)A_{\alpha}\otimes_{i=2}^{k}\left(E_{i}^{\beta_i}\ket{\widetilde{\phi
_{B_i}^{\gamma_i}}}\right)\nonumber\\ \eeqn This allows us to
rewrite (\ref{fidisom}) as \beqn
&F_e\left(\tens{i}\Psi_{(AB)_i},\mathcal{A}\circ\tens{i}\varepsilon_{i}\right)=&\\\nonumber&=
\sum_{\alpha,\beta_2 \beta_3 ... \beta_k}\sum_{\beta_1}\Big|
\sum_{\gamma_1}\bra{\widetilde{\phi
_{B_1}^{\gamma_1}}}A_{\alpha,\beta_1 \beta_2 ...
\beta_k}^{(1)}E_{1}^{\beta_1}{\alpha}\ket{\widetilde{\phi
_{B_1}^{\gamma_1}}}\Big|^2&\\\nonumber
&=F_e(\Psi_{(AB)_1},\mathcal{A}_1\circ \varepsilon_1)&\eeqn with
some channel $\mathcal{A}_1$ defined by its Kraus decomposition
\beqn\label{channelA1}\mathcal{A}_1(\cdot)=\sum_{\theta, \kappa_1
\kappa_2 ... \kappa_k}A_{\theta, \kappa_1 \kappa_2 ...
\kappa_k}^{(1)}(\cdot)A_{\theta, \kappa_1 \kappa_2 ...
\kappa_k}^{(1)\dagger}. \eeqn The above due to (\ref{reliable}) is
still high and by the Lemma 3 we can find such $W_1$ that \beqn
F_e(\Psi_{(AB)_1},\mathcal{A}_1\circ W_1)>1-2\eta. \eeqn Now
bearing in mind (\ref{krausA1}) -- (\ref{channelA1}) we conclude
that the latter can be written as \beqn
F_e\left(\tens{i}\Psi_{(AB)_i},\mathcal{A}\circ
W_{1}\otimes\left(\otimes_{i=2}^{k}\varepsilon_{i}\right)\right)>1-2\eta.\eeqn
Now one can apply the trick with defining, by partial inner
product, new channel $\mathcal{A}_2$ to the second system $B_2$
and arrive at possibility of isometric encodings on that system.
Further repeating analogous steps until one reaches $k$--th system
gives us \beqn
F_e\left(\tens{i}\Psi_{(AB)_i},\mathcal{A}\circ\tens{i}W_{i}\right)>1-2^
k\eta.\eeqn This proofs the existence of partial isometries as
good encodings. This with the aid of the fact that we still use
the same source concludes the proof.$\blacksquare$\liniaa
  \liniaa In general the isometry
can be trace--decreasing, however, as pointed in \cite{BaKniNie00}
it can be embedded in trace--preserving map without loss of
fidelity. This make it useful as an proper encoding.

\subsection{Entanglement capacities without encodings: MAC and k--UC}

The authors of \cite{BaSmoTer98} have shown that introducing
encoding to the coding scheme is not necessary to get the proper
definition of channel capacity for entanglement transmission of
SUC. We show that with additional constraint it is also the case
for MAC and $k$-UC (in the setting one sender--one receiver). One
must stress that we prove it for entanglement transmission without
assuming QAEP property of the sources. \liniaa The idea is to show
that we can get rid of encodings on the Alicias' side if Bobbys
perform an additional decoding operation. Here is the motivation
(we shall operate on the definition of capacity for entanglement
transmission). Suppose $\left(\tens{i}\ket{\phi_i^{(n)}}\right)$
is a purification of $\left(\tens{i}\varrho_i^{(n)}\right)$ which
Alicias are supplied with. Performing encoding by them means
adjoining environment in a standard state and unitary acting on
the composed system. As the environments are just an additional
subsystems which are not sent over the channel they can measure
them. This results with some probability (dependent on the result
of the measurement) in different pure states
$\left(\tens{i}\ket{\psi_i^{l(n)}}\right)$ on Alicias side.
However each Alicia have access only to a part of a resulting
state, the other is a reference system which we assume to be out
of control of any parties taking part in communication. The aim is
to show that if they use
$\left(\tens{i}\ket{\psi_i^{l(n)}}\right)$ as an input they can
also achieve high entanglement fidelity as it was for original
$\left(\tens{i}\ket{\phi_i^{(n)}}\right)$ if additional (local)
operation will be performed on Bobs' side. It will remain to show
that entropy rate of a new source is close to that of the old one.
\liniaa We start with the lemmas.
\begin{lemma}
\label{1}\cite{BaKniNie00} If
$|\inner{\phi_1}{\psi}|^{2}>1-\eta_1$ and
$|\inner{\phi_2}{\psi}|^{2}>1-\eta_2$ then
$|\inner{\phi_1}{\phi_2}|^2>1-\eta_1-\eta_2$ for normalized
$\phi_i$. \label{1}
\end{lemma}
\begin{lemma} \cite{BaSmoTer98}\label{2}
 Given a density matrix satisfying \beqn
\melement{\phi}{\varrho}>1-\epsilon \eeqn for some state $\phi$ we
have (i) \beqn \label{wlasna}\lambda_{max}>1-\epsilon \eeqn and
(ii) \beqn\label{produkt}
|\inner{\psi_{max}}{\phi}|^{2}>1-2\epsilon \eeqn where
$\lambda_{max},\psi_{max}$ are the largest eigenvalue and
corresponding eigenstate respectively.
\end{lemma}
The central element of the technique to prove the upcoming Theorem
is the subsequent Lemma 7 which is a subtle generalization of the
corresponding lemma from \cite{BaSmoTer98}.
\begin{lemma} \label{lemmapuryfikacja} Given a density matrix $\varrho$ in
Hilbert space $\ch_{\bold{AB}}=\tens{i}\ch
_{(AB)_i}(\ch_{(AB)_i}=\ch_{A_i}\otimes\ch_{B_i}) $ satisfying a
condition
$(\tens{i}\bra{\phi_i}){\varrho}|(\tens{i}\ket{\phi_i})>1-\epsilon$
there exists a purification $\ket{\Psi}=\tens{i}\ket{\Psi_i}$ of
$\tens{i}\rho_i \equiv\tens{i}\tr _{\bold{AB}\setminus A_i}
\varrho$ into Hilbert space $\ch_{\bold{ABC}}=\tens{i}\ch
_{(ABC)_i}(\ch_{(ABC)_i}=\ch_{A_i}\otimes\ch_{B_i}\otimes\ch_{C_i})$
such that \beq \bra{\Psi}(\varrho\otimes (\tens{i}0
^{C_i}))\ket{\Psi}
>1-O(\epsilon)\;,
\eeq where we can take $O(\epsilon)=(2k+4)\epsilon$.
\end{lemma}
{\it Proof:} We can write \beqn \varrho\otimes(\tens{i}
0^{C_i})=\lambda_{\rm max} \proj{\eig}\otimes(\tens{i}0^{C_i})
+\\\nonumber (1-\lambda_{\rm max})
\varrho'\otimes(\tens{i}0^{C_i})\eeqn with $\lambda_{max}$ largest
eigenvalue of $\varrho$ and $\phi_{max}$ corresponding
eigenvector, and take purifications in the form \beqn
&&\hspace{-8mm}\ket{\Psi_i}= \sqrt{\lambda^{(i)}_{\rm max}}
\ket{\phi _{max}^{(i)}}\otimes\ket{0^{C_i}}\no &&+
\sqrt{1-\lambda^{(i)}_{\rm max}}\sum_{k=1}^{\dim \ch_{A_i}}
\sqrt{\mu_k} \ket{\phi_k ^{AB}} \otimes \ket{k^{C_i}} \eeqn Now
\beqn &&\hspace{-9mm}\melement{\Psi}{(\varrho\otimes
(\tens{i}0^{C_i} )} =\no&&=(\tens{i}\bra{\Psi_i})\varrho\otimes
(\tens{i}0^{C_i} ) )(\tens{i}\ket{\Psi_i})=\no&&=
\lambda_{max}|(\tens{i}\bra{\Psi_i})}|{\psi_{max}\otimes
(\tens{i}0^{C_i} )|^{2}=\no&&=
\lambda_{max}\Pi_{i=1}^{k}\lambda^{(i)}_{max}|(\tens{i}\bra{\phi
_{max}^{(i)}})\ket{\psi_{max}}|^{2}\ge\no&&\hspace{+45mm}\ge
1-O(\epsilon).\eeqn which concludes the proof. The inequality
follows from lemmas 1 and 6.$\blacksquare$ \linia The preceding
lemma is crucial for proving the main theorem of this section
which is the following.
\begin{theorem}
For a given reliable protocol
$P=\{\mathcal{D}^{(n)},\tens{i}\varepsilon_{i}^{(n)}\}$ for MAC
and $k$--UC there always exists a reliable protocol
$\tilde{P}=\{(\tens{i}
\tilde{\mathcal{D}}_{i}^{(n)})\mathcal{D}^{(n)},\mathbb{I}\}$
allowing for entanglement transmission with the same rate.
\end{theorem}
\proof Assuming reliable transmission and taking
$\tens{i}\ket{\Psi_i^{(n)}}$ purifying $\tens{i}\tr
_{\bold{AB}\setminus A_i} \varrho_{\bold{AB}} ^{out(n)}$ we have
by Lemma 7 \beq \bra{\Psi^{(n)}}\varrho^{\rm
out(n)}_{\bold{AB}}\otimes (\tens{i}0^{C_i})\ket{\Psi^{(n)}}
>1-O(\epsilon),\eeq where $\varrho^{\rm out(n)}_{\bold{AB}}$ is the
density matrix after performing protocol.
 One can see that  \beq \label{unitary}
\ket{\Psi^{(n)}}=\mathbb{I}_{\bold{A}}\otimes
(\tens{i}U^{\dagger}_{(BC)_i})\ket{\psi_0^{l(n)}}\equiv\mathcal{U}
\ket{\psi_0^{(n)}},\eeq where $\ket{\psi_0
^{l(n)}}\equiv(\tens{i}\ket{\psi_i ^{l(n)}})\otimes
(\tens{i}\ket{0^{C_i}})$ and $U$ is unitary, as both
$\ket{\Psi^{(n)}}$ and $\ket{\psi_0^{l(n)}}$ are purifications of
the same state. We get \beqn
\melement{\psi_{0}^{(n)}}{\mathcal{U}\varrho_{\bold{0}}^{out}
\mathcal{U}^{\dag} }\ge 1-O(\epsilon), \eeqn with substitution
$\varrho_{\bold{0}}^{out(n)}=\varrho_{\bold{AB}}^{out(n)}\otimes(\tens{i}0^{C_i})$.
We see that if we add remaining terms with environment in other
states to get the full trace we obtain \beq
(\tens{i}\bra{\psi_i^{(n)}}){\tr_{\bold{C}}
\mathcal{U}\varrho_0^{\rm out(n)}
\mathcal{U}^\dag}(\tens{i}\ket{\psi_i^{(n)}}) \ge 1-O(\epsilon).
\eeq So replacing input $\left(\tens{i}\ket{\phi_i^{(n)}}\right)$
with $\left(\tens{i}\ket{\psi_i^{l(n)}}\right)$ and adding
additional decoding operation $\widetilde\cd$ -- appending $k$
environments, rotating the whole state locally and tracing out
environments -- allowed us to transmit a given source reliably.
Structure of an additional operation is essential here. One can
see from (\ref{unitary}) that it must be alike coding operation.
This is what makes the technique used above useful only in the
case of $k$--UC (in the setting one sender--one receiver) and MAC.
It is interesting that although Bobby can perform in case of MAC
global decoding it suffices to perform local operations to achieve
reliable transmission. Previous considerations fail in case of
general broadcast channel (only in the case of broadcast channels
for which product coding suffices we can apply our theorem). To
conclude the proof it remains to show that entropy of new sources
producing $\varrho'^{(n)}\equiv
\tr_{\bold{A}}(\tens{i}\psi_i^{l(n)})$ is close to the old one
producing $\varrho^{(n)} \equiv
\tr_{\bold{A}}(\tens{i}\phi_i^{(n)})$.
  Following lemma, which is a simple generalization of lemma from
  \cite{BaSmoTer98},
  will be deciding.
\begin{lemma}
\label{} For a given pure state $\ket{\phi}\equiv\tens{i}\ket{\phi
_i}$ and density matrix $\varrho$ in Hilbert space
$\ch=\tens{i}\ch _{(AB)_i}$ with $\melement{\phi}{\varrho} \ge
1-\epsilon$ and $\epsilon < \frac{1}{72}$ we have \beq
|S(\tr_{\bold{A}} \phi) - S(\tr_{\bold{A}} \;\varrho)| \le 2
\sqrt{2 \epsilon} \log \dim \ch_{\bold{B}} + 2.\eeq
\end{lemma} We have by the lemma and
high entanglement fidelity assumption for
$\left(\tens{i}\ket{\phi_i^{(n)}}\right)$ \beqn |S(\tr_{\bold{B}}
(\tens{i}\phi_i^{(n)}) - S(\tr_{\bold{B}}
\varrho_{\bold{AB}}^{out(n)})| \le\no 2 \sqrt{2 \epsilon} \log
\dim \ch_{\bold{A}}^{(n)} + 2 \eeqn for $\epsilon < \frac{1}{72}$.
Following equalities hold $\tr_{\bold{B}}\ \varrho^{\rm
out}_{\bold{AB}} = \tr_{\bold{B}}
\left(\tens{i}\psi_i^{(n)}\right)$, $S\left(\tr_{\bold{B}}
\left(\tens{i}\psi_i^{(n)}\right)\right)=S\left(\tr_{\bold{A}}
\left(\tens{i}\psi_i^{(n)}\right)\right)\equiv
S\left(\varrho'^{(n)}\right)$, $S\left(\tr_{\bold{B}}
\left(\tens{i}\phi_i^{(n)}\right)\right)$\newline$=S\left(\tr_{\bold{A}}
\left(\tens{i}\phi_i^{(n)}\right)\right)\equiv S(\varrho^{(n)})$
which immediately implies \beq
|S(\varrho^{(n)})-S(\varrho'^{(n)})| \le 2 \sqrt{2 \epsilon} \log
\dim \ch_A ^{(n)}+ 2. \eeq The latter means that in the limit of
large $n$ we achieve the same transmission rate for the new
source. The above reasoning holds for each "subtransmission", \tzn
each local entropy, which means that information is localized
without changes. \liniaa However, we must stress again we are not
able to show that new source is QAEP if the original one was. Most
probably it is not possible in general, \tzn encodings is
necessary to preserve QAEP (besides the trivial case when input
density matrix is almost maximally chaotic on channel input
space).

\section{Forward classical communication does not improve capacity regions}

Now we turn to the case when quantum transmission is supplemented
by classical noiseless forward channel. It is well known fact that
classical support does not increase capacity \cite{BeDiSmoWoo96},
\cite{BaKniNie00} of SUC. We show this is the case for all quantum
channels considered in the paper. Our strategy will be to
construct a reliable zero--way protocol from reliable one--way
protocol without changing the rate of transmission. To this aim
consider a set of protocols, indexed by $j_i$ (which in fact is an
multiindex) representing classical messages sent by Alicias,
$\mathcal{P} ^{\rightarrow}_{j_i}=\{
\otimes_{i=1}^{m}\cd_i^{j_i(n)}, \tens{i}\varepsilon_i^{j_i(n)}
\}$ with $\left(\tens{i}\varepsilon_i^{j_i(n)}\right)$ summing
over set of messages to a trace--preserving operation and each
$\left(\otimes_{i=1}^{m}\cd_i^{j_i(n)}\right)$ trace--preserving.
As mentioned we have for the protocol $\sum
_{j_i}F_e\left(\tens{i}\left(\otimes_{\alpha=1}^{l_i}\Psi_i^{\alpha(n)}\right),
\otimes_{i=1}^{m}\cd_i^{j_i(n)}\ca\tens{i}\varepsilon_i^{j_i(n)}\right)>1-\eta$
which means that for one value $j_i\equiv j$ we have
$F_e\left(\tens{i}\left(\otimes_{\alpha=1}^{l_i}\Psi_i^{\alpha(n)}\right),
\otimes_{i=1}^{m}\cd_i^{j(n)}\ca\tens{i}\varepsilon_i^{j(n)}/\right.$\newline$\left.
\tr \;\left(\tens{i}\varepsilon_{i}^{j(n)}\right)
\left(\tens{i}\left(\otimes_{\alpha=1}^{l_i}\varrho_i^{\alpha(n)}\right)\right)\right)>1-\eta$
which by Theorem 2 implies existence of a reliable protocol using
extendable isometries as an encoding. This means construction of a
reliable zero--way protocol without changing the rate which shows
uselessness of classical forward communication in quantum
information transmission.

\section{Discussion}

We have considered general multiparty scenario of quantum
channels. We have considered the capacity of coherent quantum
transfer in this scenario. We have defined capacities under
minimal subspace transmission and entanglement transmission and
have shown that, like in bipartite scenario,  the transmission do
coincide. The alternative notions of fidelities for both scenario
have also been considered and shown to be equivalent. We have also
proven that in multiparty scenario forward classical communication
does not help. This was achieved by generalization of bipartite
theorem on sufficiency of isometric encoding. The result proves
optimality of recently derived by other authors zero--way capacity
regions also for one--way scenario. We have also considered
multiple access channel and $k$--user channel separately and shown
that entanglement transmission capacity can be achieved also
without encoding. This result however does not seem to be true for
broadcast channel and in cases when one assumes QAEP sources.

The results of the present paper can be applied to get simple
capacity regions for quantum broadcast and $k$--user channels, but
this will be considered elsewhere \cite{Prep}.

\acknowledgments We thank Michal Horodecki for discussions. MD
also acknowledges discussions with Remigiusz Augusiak. This work
was prepared under the EC IP project SCALA and the Polish Ministry
of Scientific Research and Information Technology project
(solicited) no. PBZ-MIN-008/P03/2003 .

\section{Appendix}

\subsection{Quantum Asymptotic Equipartition Property (QAEP)}

Here we recall definitions for quantum analogs of typical subspace
and equipartition property. Classical formulas are easily adopted
to the quantum case and one has the
following\begin{definition}\cite{BaKniNie00} We define
$\epsilon$-- typical subspace of a $n$--block $\varrho^{(n)}$
produced by a quantum source $\Sigma$ on a Hilbert space $H$ to be
the subspace $T_{\epsilon}^{(n)}$ of $H^{\otimes n}$ spanned by
the eigenvectors $\ket{\lambda}$ of $\varrho^{(n)}$ with
eigenvalues $\lambda$ satisfying\beqn
2^{-n(S(\Sigma)+\epsilon)}\le \lambda\le
2^{-n(S(\Sigma)-\epsilon)}\eeqn
\end{definition}
\begin{definition}\cite{BaKniNie00} We say a quantum source
$\Sigma$ producing $n$--block material $\varrho^{(n)}$ satisfies
QAEP iff for any positive $\epsilon$ and $\delta$ in the limit of
large $n$ the $\epsilon$--typical subspace of $\varrho^{(n)}$
satisfies \beqn \tr \Lambda
^{(n)}\varrho^{(n)}\Lambda^{(n)}>1-\delta\eeqn where $\Lambda
^{(n)}$ denotes the projection onto $T_{\epsilon}^{(n)}$.
\end{definition}We can think about $\epsilon$--typical subspace as
a small set containing almost all probability. Obviously a tensor
product of QAEP sources is also QAEP.

 \subsection{Proof of Eq.(\ref{fidelity})}

Consider global entanglement fidelity $F_e
\left(\tens{i}\left(\otimes_{\alpha=1}^{i}\Psi
_{i}^{\alpha(n)}\right),\ca ^{(n)}\right)\ge 1-\eta$ with $\eta
=\sum_{l=1}^{L}\eta _{l}$. By convexity of entanglement fidelity
in the input operator we have the following \beq \label{fid}
1-\eta\le (1-\gamma)\Pi _{l=1}^{L}\alpha
_{l}+\Pi_{l=1}^{L}(1-\alpha _{l}),\eeq where $\gamma$ describes
imperfection of global pure state transmission of vectors from the
subspace after having removed dimensions (\tzn $F_{s}\ge
1-\gamma$). One can verify that \beq\label{nierownosc}
\Pi_{l=1}^{L}\alpha _{l} +\Pi_{l=1}^{L}(1-\alpha _{l})\le 1 \eeq
as for $k$ it is trivially true and multiplying of each component
just lowers the number. With the aid of the above and results of
IID we immediately conclude the bound for {\it local} pure--state
fidelity \beq F_s^{(l)}\ge 1-\displaystyle\frac{\sum_{l=1}^{L}\eta
_{l}}{\Pi_{l=1}^{L}\alpha _{l}} .\eeq


\begin{thebibliography}{}
\bibitem{Schu95} B. Schumacher, Phys. Rev. A {\bf 51}, 2738 (1995).
\bibitem{Schu96} B. Schumacher, Phys. Rev. A {\bf 54}, 2614 (1996).
\bibitem{BeDiSmoWoo96}C.H.~Bennett, D.P.~DiVincenzo, J.A.~Smolin, W.K.~Wootters,
 Phys. Rev. A. {\bf 54}, 3824 (1996).
\bibitem{BaNieSchu98} H. Barnum, M.A. Nielsen and B. Schumacher, Phys. Rev. A {\bf 57}, 4153
(1998).
\bibitem{BaKniNie00} H. Barnum, E. Knill, M.A. Nielsen, IEEE Trans. Inf. Th. {\bf
46}, 19 (2000).
\bibitem{Llo97} S. Lloyd, Phys. Rev. A. {\bf 55}, 1613 (1997).
\bibitem{BeDiSmo97}C.H. Bennett, D.P. DiVincenzo and J.A. Smolin,
Phys. Rev. Lett. {\bf 78}, 3217 (1997).
\bibitem{BaSmoTer98} H.Barnum, J. A. Smolin, B. Terhal, Phys. Rev.
A {\bf 58}, 3496 (1998).
\bibitem{HoHoHo00} M. Horodecki, P. Horodecki, R. Horodecki, Phys.
Rev. Lett. {\bf 85}, 433 (2000).
\bibitem{DevetakWinter}I. Devetak, A. Winter, Phys. Rev. Lett., {\bf 93}, 080501,
(2004); I. Devetak, A. Winter, Proc. R. Soc. Lond. A {\bf 461},
207 (2005).
\bibitem{De05} I. Devetak, IEEE Trans. Inf. Th. {\bf IT-55}, 44 (2005).
\bibitem{KreWer03} D. Kretschmann, R. Werner, New J. Phys.
\textbf{6}, 26 (2004).
\bibitem{Ho96}A.S. Holevo, Report No. quant-ph/9611023.
\bibitem{EntanglementAssistedCapacity} C. H. Bennett, P. W. Shor, J. A. Smolin, A. V.
Thapliyal,
Phys. Rev. Lett. {\bf 83}, 3081 (1999).
\bibitem{Sho03} P. W. Shor, Report No. quant-ph/0305035.
\bibitem{BeDeShoSmo04} C.H. Bennett, I. Devetak, P.W. Shor, J.A.
Smolin, Report No. quant-ph/0406086v1.
\bibitem{BeShoScience}C. H. Bennett, P. Shor, Science
{\bf 303}, 1784 (2004).
\bibitem{HoHoHoOpp05}
K. Horodecki, M. Horodecki, P. Horodecki, J. Oppenheim, Phys. Rev.
Lett. {\bf 94}, 160502 (2005).
\bibitem{Winter2001}
A. Winter, IEEE Trans. Info. Th. {\bf IT-47},  3059 (2001).
\bibitem{DuHoCi04}
W. D\" ur, J. I. Cirac, and P. Horodecki, Phys. Rev. Lett. {\bf
93}, 020503 (2004).
\bibitem{superactivation}
P. W. Shor, J. A. Smolin, and A. V. Thapliyal, Phys. Rev. Lett.
{\bf 90}, 107901 (2003).
\bibitem{Nature2005}
M. Horodecki, J. Oppenheim, A. Winter, Nature {\bf 436}, 673
(2005).
\bibitem{NatureExtension}
M. Horodecki, J. Oppenheim, A. Winter, quant-ph/0512247.
\bibitem{YaDeHay05} J. Yard, I. Devetak, P. Hayden, Report No.
quant-ph/0501045.
 \bibitem{Ho03} P. Horodecki, Centr. Europ. J. Phys. {\bf 4}, 695
(2003).
\bibitem{Prep}
M. Demianowicz, P. Horodecki, "Capacity regions for multiparty
quantum channels", in preparation.
\end{thebibliography}
\end{document}